\documentclass[10pt,aps,notitlepage,prb,twocolumn,showpacs,preprintnumbers,amsmath,amssymb,floatfix]{revtex4-2}

\usepackage{graphicx}
\usepackage{bm}
\usepackage[colorlinks,allcolors=blue]{hyperref}
\usepackage{braket}

\begin{document}
\title{Quantum Hopfion rings in the cluster mean-field approximation}
\author{Vladyslav M. Kuchkin}
\email{vladyslav.kuchkin@uni.lu}
\affiliation{Department of Physics and Materials Science, University of Luxembourg, 162A~Avenue de la Faiencerie, L-1511 Luxembourg, Grand Duchy of Luxembourg}

\author{Thomas L. Schmidt}
\affiliation{Department of Physics and Materials Science, University of Luxembourg, 162A~Avenue de la Faiencerie, L-1511 Luxembourg, Grand Duchy of Luxembourg}

\makeatother

\begin{abstract}
We study the quantum properties of two- and three-dimensional spin textures -- $k\pi$-skyrmions and hopfion rings -- within the cluster mean-field approximation (CMFA).
By combining the CMFA with a symmetrization procedure, we achieve two key advances: the accurate computation of quantum fluctuations in large spin textures and reliable access to metastable states. 
These challenges are generally insurmountable using standard methods, which are severely limited by the curse of dimensionality and typically restricted to ground-state properties.
Exploiting the cylindrical symmetry of the studied magnetic configurations, we construct one-dimensional chain-like clusters that can be efficiently simulated using the density matrix renormalization group method, while inter-cluster interactions are treated at the mean-field level.
The resulting spatial profiles of quantum features such as the local variation of the magnetization length in hopfion rings reveal limitations of the classical micromagnetic model and indicate the necessity of its extension. 
We demonstrate that the recently proposed regularized micromagnetic equation provides a suitable framework for this purpose.
\end{abstract}

\maketitle
\noindent
Due to their stability and mobility, topologically nontrivial magnetic spin textures are considered as attractive for the next generation of computing devices -- from classical~\cite{Tokura2020,Song2020} to quantum~\cite{Psaroudaki2021,Xia2023,Petrovi2025,Chudnovsky2025}.
Skyrmions and hopfions are the most prominent examples of such textures and are characterized by the integer indices,
\begin{align}
    \!\!\!Q &= \dfrac{1}{4\pi}\int_{\Omega} \mathrm{d}x\mathrm{d}y\,F_{z}, & 
    H &=-\dfrac{1}{16\pi^{2}}\int_{\Omega} \mathrm{d}x\mathrm{d}y\mathrm{d}z\,\mathbf{A}\cdot\mathbf{F},\label{Qhopf}
\end{align}
respectively, where the gyro-vector $\mathbf{F}$ has components $F_{i}=\epsilon_{ijk}\mathbf{m}\cdot\left(\partial_{j}\mathbf{m}\times\partial_{k}\mathbf{m}\right)$ and vector potential $\mathbf{A}$ satisfies the equation $\mathbf{F}=\nabla\times\mathbf{A}$.
The magnetization unit vector $\mathbf{m}(\mathbf{r})$ is defined inside the magnetic medium with volume $\Omega \subset \mathbb{R}^{2}$ for the 2D case and $\Omega \subset \mathbb{R}^{3}$ for the 3D case. On the boundary $\partial \Omega$, $\mathbf{m}$ is constant. 
The values of $Q$ and $H$ are typically preserved under weak perturbations, e.g., pulses of the field, current~\cite{Juge2019}, laser~\cite{Tengdin2022}, or thermal fluctuations~\cite{Hirosawa2022}, and lead to the topological stability of the magnetic state.
The creation and annihilation of such states, though possible at stronger perturbations~\cite{Lemesh2018,ViasBostrm2022}, require the appearance of Bloch points~\cite{Feldtkeller1965, Doring1968} and are typically suppressed by energy barriers~\cite{Chen2026_2}.
In this work, we are particularly interested in symmetric states such as skyrmion bags~\cite{Foster2019,Kern2025} and hopfion rings~\cite{Zheng2023,Chen2026_1}, which have been experimentally reported recently. 
Materials with competing Heisenberg exchange and Dzyaloshinskii-Moriya interactions (DMI)~\cite{Muhlbauer2009,Lancaster2016} seem to be ideal for hosting a plethora of topological states in 2D~\cite{Bogdanov_89, Bogdanov_1994, Rybakov2019, Kuchkin2020} and 3D~\cite{Voinescu2020,Kuchkin2023, Kuchkin2022}.

\begin{figure*}[tb!]
\centering
\includegraphics[width=17.5cm]{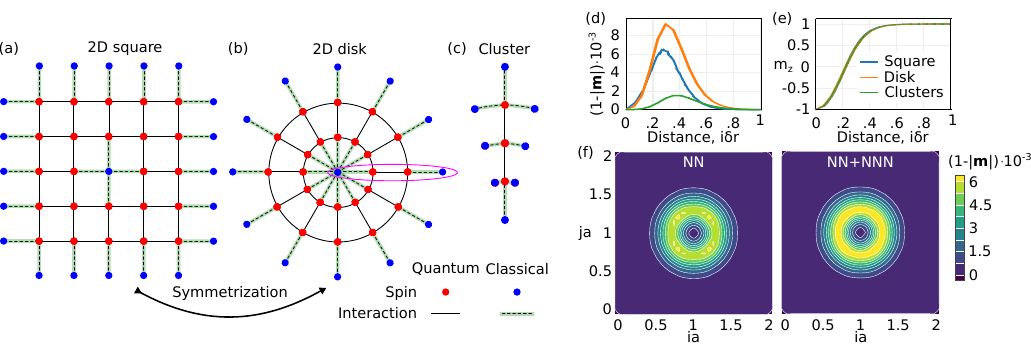}
\caption{Results of 2D DMRG and CMFA simulations for a quantum skyrmion. 
The positions of quantum (red) and classical (blue) spins and their couplings are shown for square (panel a), disk (b), and cluster (c) geometries. 
(Panel d) Change of magnetization length $|\mathbf{m}|$ as a function of the distance $r$ to the skyrmion center obtained for the geometries (a)-(c).
(Panel e) Magnetization $m_{z}$ as a function of $r$ for cases (a)-(c).
(Panel f) Distribution of the magnetization length $|\mathbf{m}|$ for the cases when only nearest-neighbor (NN) spin interactions are taken into account (left) and with inclusion of next-nearest-neighbor (NNN) spin interactions (right).
} 
\label{Fig1}
\end{figure*}

Our main goal in this Letter is to study the quantum features of these topological states. For this purpose, we develop an approach which relies heavily on the symmetry of the magnetic spin texture and which allows us to simplify the model Hamiltonian considerably.  
One of the main challenges in solving quantum spin models is the so-called curse of dimensionality, which limits our computational resources even to store the quantum wave function associated with a given magnetic state. 
The mean-field approximation (MFA), which corresponds to assuming a product state wave function, makes it possible to study large systems but only in the classical regime.
In this work, we are interested in the intermediate regime between classical and quantum spin textures, which is accessible within the cluster mean-field approximation (CMFA)~\cite{Yamamoto2009}. It was first introduced and then widely used by Bethe~\cite{Bethe1935}, Peierls~\cite{Peierls1936}, and Weiss~\cite{Weiss1948} for studying magnetic properties at finite temperatures.
Following this approach, we split a large quantum spin system into smaller clusters, each of which is treated quantum-mechanically, while inter-cluster interactions are treated in the MFA.
As one might expect, in the CMFA the final results can vary with how the clusters are chosen, so a judicious choice of the cluster geometry, reflecting the symmetries of the targeted states, is essential. 
The aim of our approach is to capture quantum features that are preserved in large systems near the thermodynamic limit, where one can expect the state to be almost classical. In this case, one expects the symmetry of the classical magnetic texture to be reflected in the quantum mechanical wave function.
Based on that, we start by expressing the quantum Heisenberg model in cylindrical coordinates and choose clusters in the form of 1D chains along the radial direction, thus neglecting entanglement in the other two spatial directions [see Fig.~\ref{Fig1}(a)-(c)]. 
The justification for such a choice is the axially symmetric distribution of entanglement, magnetization length, and von Neumann entropy reported earlier for 2D skyrmions~\cite{Sotnikov2021,Haller2022,Mazurenko2023,Salvati2024,Lik2026}.

Another key advantage of the CMFA is its ability to access excited (or metastable) states. 
This feature is particularly valuable for topologically non-trivial textures such as isolated skyrmions and hopfion rings, which typically do not appear as ground states.
For benchmarking our results for excited states, we compare them with results for smaller systems, where we use specific boundary conditions to turn the excited state into ground states, enabling their study with the density matrix renormalization group (DMRG) method~\cite{White1992}.
For larger systems with $k\pi$-skyrmions and hopfion rings, we rely exclusively on the CMFA.
This approach seems particularly attractive for theoretical study, as it provides an upper bound on the quantum fluctuations and entanglement that can occur in the system~\cite{Rzsa2025,Nishimura2025}.
For a given magnetic state, denoting its energy in the quantum model, CMFA, and MFA as $E_\mathrm{q}$, $E_\mathrm{CMFA}$, and $E_\mathrm{MFA}$, respectively, we have $E_\mathrm{q}\leq E_\mathrm{CMFA} \leq E_\mathrm{MFA}$. 
The latter inequalities can be obtained by neglecting entanglement and hold generically for particles with any spin $s$ at low temperatures.
For the sake of generality, we retain the spin-$s$ notation throughout the formulas below, while performing all numerical simulations for the most quantum case of $s=1/2$.
This choice is motivated by the fact that larger spin values are expected to exhibit more classical behavior~\cite{Millard1971, Lieb1973}.

\textit{Classical and quantum models on a cubic lattice} -- The chiral magnet Hamiltonian in the MFA is given by
\begin{align}
\!\!\!\mathcal{E}=\int_{\Omega}\mathrm{d}V\left[w_\mathrm{ex}+2\pi w_\mathrm{dmi}+4\pi^{2}w_{\mathrm{u}}\right],\label{E_dimensless}
\end{align}
where $w_\mathrm{ex}=\left(\nabla\mathbf{m}\right)^{2}/2$ and $w_\mathrm{dmi}=\mathbf{m}\cdot\nabla\times\mathbf{m}$ are the exchange and bulk DMI energy densities, respectively. The potential term $w_{\mathrm{u}}=h\left(1-m_{z}\right)+u\left(1-m_{z}^{2}\right)$ accounts for an external magnetic field ($h$) and uniaxial anisotropy ($u$), which both act along the $z$-axis.
The dimensionless form of Eq.~\eqref{E_dimensless} presented is obtained in a standard way~\cite{Bogdanov_89,Bogdanov_1994}, and details are provided in Supplemental Material I.

The atomistic (discrete) version of Eq.~\eqref{E_dimensless} is obtained by replacing $\mathbf{m}\to \mathbf{s}/s$ and is written as,
\begin{align}
    E &= aJ\sum_{ij}\left(1-\dfrac{\mathbf{s}_{i}\cdot\mathbf{s}_{j}}{s^2}\right) - 2\pi a^{2}D\hat{\mathbf{r}}_{ij}\cdot\sum_{ij}\dfrac{\mathbf{s}_{i}\times\mathbf{s}_{j}}{s^2} \nonumber\\
    &+4\pi^{2} a^{3}\sum_{i}\left(h+2u\dfrac{\braket{s_{i}^{z}}}{s}\right)\left(1-\dfrac{s_{i}^{z}}{s}\right),\label{XXZ}
\end{align}
where $a$ is a cubic lattice constant,  $\mathbf{s}_{i}=\left(s^{x}_{i},s^{y}_{i},s^{z}_{i}\right)$ is a spin operator at site $i$ for a spin-$s$ particle, and $J$ and $D$ are dimensionless exchange and DMI constants.
The DMI is of bulk type, so $\hat{\mathbf{r}}_{ij}=(\mathbf{r}_{i}-\mathbf{r}_{j})/|\mathbf{r}_{i}-\mathbf{r}_{j}|$.
Below, we consider the case where the summation runs over the nearest-neighbor spins, corresponding to $J=D=1$, and the case where next-nearest-neighbor spins also interact with strengths $J_{1}=4/3$, $J_{2} = -1/12$, $D_{1}=4/3$, and $D_{2}=-1/6$. These values correspond to the second- and the fourth-order finite-difference approximation~\cite{Donahue1997} of the derivatives in Eq.~\eqref{E_dimensless}.
The anisotropy term is included as an interaction with the effective field, which is defined by the on-site spin expectation values. We set $u=0$ unless stated otherwise.

Using the DMRG method, we study the 2D skyrmion excitation~\cite{Romming2013,Takashima2016,Petrovi2025} in the model \eqref{XXZ} at $h=0.85$, corresponding to a ferromagnetic ground state. A discussion of convergence is provided in the Supplemental Material III.
The skyrmion core and the far-field spins are fixed classically, while all other spins are treated quantum-mechanically [see Fig.~\ref{Fig1}(a)].
The magnetization length $|\mathbf{m}| = |\! \braket{\mathbf{s}_{ij}}\!/s|$, which must be equal to one for classical spins but can be less than one for quantum spins, is shown in panels (d) and (e), while the $m_{z}$ component of the skyrmion profile is shown in panel (e). 
The model with only nearest-neighbor interactions still exhibits a weakly broken axial symmetry, particularly visible on the square diagonals, which we attribute to lattice effects.
In the model with both nearest- and the next-nearest-neighbor interactions, this lattice effect is suppressed, and the distribution of the magnetization length becomes axially symmetric with high accuracy.
This is an important aspect because in the thermodynamic limit the state properties must be free of lattice effects, as the classical continuum model \eqref{E_dimensless} predicts.
This axial symmetry is the main motivation for considering the quantum model in cylindrical coordinates, which we provide below.
From this moment on, we do not aim for quantitative agreement between these models because matching energy densities in the cubic and cylindrical geometries requires fine-tuning the number of nodes in different spatial directions, and even after doing so, the match will be only approximate due to the incommensurability of these geometries.
Instead, we are interested in capturing symmetric quantum features, which are naturally represented by the Hamiltonian expressed in cylindrical coordinates.

\textit{Quantum model in cylindrical coordinates} -- We introduce cylindrical coordinates in continuum Hamiltonian \eqref{E_dimensless}, and approximate corresponding derivatives by finite differences: $\partial_{r}\mathbf{m}\to \left(\mathbf{m}_{i+1,jk}-\mathbf{m}_{ijk}\right)/\delta r$ and $\partial_{\phi}\mathbf{m}\to \left(\mathbf{m}_{i,j+1,k}-\mathbf{m}_{ijk}\right)/\delta\phi$.
The positions of spins in the discrete Hamiltonian are shown in Fig.~\ref{Fig1}(b).
For convenience, we use a notation where an index $ijk$ denotes a point in a 3D space in the cylindrical basis, i.e., $\mathbf{m}_{ijk}=\mathbf{m}(i \delta r\mathbf{e}_{r}+j \delta \phi\mathbf{e}_{\phi} +k \delta z\mathbf{e}_{z})$. For 2D skyrmions, we omit the index $k$.
The dimensionless variables $\delta r$, $\delta \phi$, and $\delta z$ correspond to the distances between the particles in the radial, azimuthal, and $z$-axis directions, respectively.
The resulting Heisenberg Hamiltonian is analogous to the one given in Eq.~\eqref{XXZ}, but due to the coordinate transformation now explicitly depends on spatial variables (see Supplemental Material II).

\begin{figure}[tb!]
\centering
\includegraphics[width=8cm]{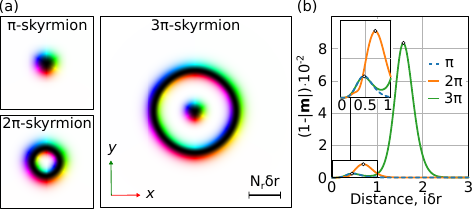}
\caption{~\small Quantum $k\pi$-Skyrmions. 
\emph{Panel (a):} Examples of $k\pi$-skyrmions stabilized in the system of size $48$ spins with $\delta r=1/16$ and $\delta \phi = 2\pi/16$. 
Spin directions are indicated by the following color code: white for up, black for down, and a red–green–blue color scheme for in-plane spin components~\cite{Savchenko2022}.
\emph{Panel (b):} Distribution of magnetization as a function of the distance to the skyrmion center obtained with the CFM approach.
Maxima points are denoted by hollow circles, and the corresponding values for $\pi$-, $2\pi$-, and $3\pi$-skyrmions are $0.0028$, $0.0085$ and $0.083$, respectively.
} 
\label{Fig2}
\end{figure}

The results of the 2D DMRG simulations for the skyrmion in cylindrical coordinates are shown in Fig.~\ref{Fig1}(d) and (e).
We consider an identical number of lattice sites in the radial ($N_{r}$) and azimuthal ($N_{\phi}$) directions, $N_{\phi}=N_{r}=16$, which leads to comparable results obtained for the model \eqref{XXZ}.
The
magnetization profile $\braket{\mathbf{s}_{ij}}/s$ of the skyrmion coincides with that obtained in the Hamiltonian on a cubic lattice (e).
However, the magnetization length shown in panel (d) does not exactly match the one obtained after solving the Hamiltonian \eqref{XXZ}.
This is expected because the Hamiltonians in both geometries will coincide only for a larger number of lattice sites.

\textit{Skyrmion in the CMFA} -- Next, we consider the CMFA for the quantum model in cylindrical coordinates.
Assuming that the entanglement is strongest along the radial direction \cite{Haller2022}, we treat interactions in the azimuthal direction in the MFA, i.e. $s_{ij}^{\alpha}s_{i,j\pm1}^{\beta}\to s_{ij}^{\alpha}\braket{s_{i,j\pm1}^{\beta}}$ (for $\alpha, \beta \in \{x,y,z\}$), while keeping those in the radial direction quantum mechanical [see Fig.~\ref{Fig1}(c)]. 
As the skyrmion has axial symmetry, the spin expectation values on the adjacent clusters are
\begin{eqnarray}
    \braket{\mathbf{s}_{i,j\pm1}}=\mathcal{R}(\pm\delta\phi)\braket{\mathbf{s}_{ij}},\label{excpetc_values}
\end{eqnarray}
where $\mathcal{R}$ is a $3\times 3$ rotation matrix about the $z$-axis, the index $j$ selects the cluster and the index $i$ numbers the individual spins in the cluster.
The boundary conditions (BC) for a skyrmion are accounted for via an interaction with classical spins with values $(0,0,-s)$ at the origin and $(0,0,s)$ on the circumference.
Taking into account Eq.~\eqref{excpetc_values}, the 2D Hamiltonian in the CMFA transforms into an effective 1D problem (see Supplemental Material II), which can be solved numerically using the DMRG method.
To satisfy the self-consistency criterion in the CMFA, we update the expectation values $\braket{s_{ijk}^{\alpha}}^{(\rm old)} \to \braket{s_{ijk}^{\alpha}}^{({\rm new})}$ after each converged DMRG simulation.

\begin{figure*}[tb!]
\centering
\includegraphics[width=\textwidth]{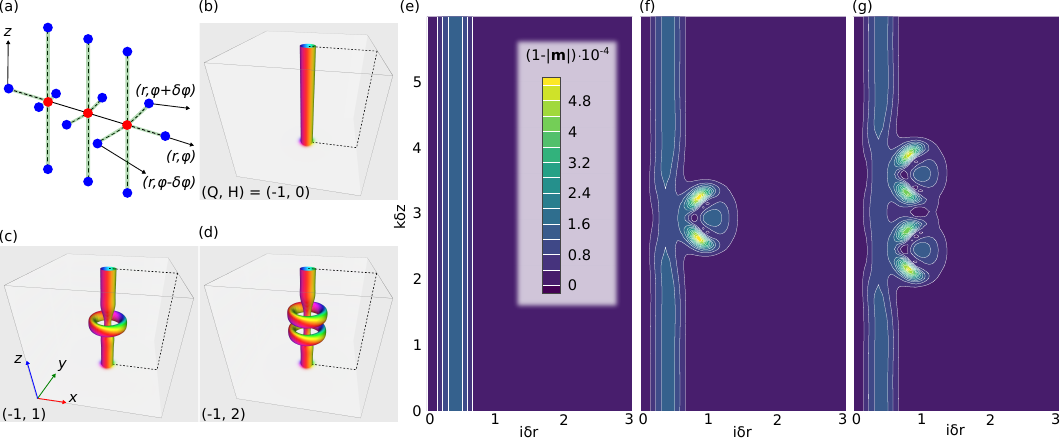}
\caption{~\small Quantum Hopfion rings. 
\emph{Panel (a):} positions of quantum (red) and classical (blue) spins and their couplings for the cluster in the cylinder geometry -- a 3D extension of the one provided in Fig.~\ref{Fig1}(c).
\emph{Panels (b), (c) and (d):} skyrmion strings with zero, one and two hopfion rings, respectively, stabilized at $h=0.4$, $u=0.3$ in the box of size $\left(96\right)^{3}$, with $a=1/16$ in the model \eqref{E_dimensless}.
\emph{Panels (e)-(g):} magnetization distribution obtained with the CFMA in the rectangular region bounded by the skyrmion string core spins and dashed lines in panels (b)-(d).
} 
\label{Fig3}
\end{figure*}

The results of the CMFA simulations are shown in Fig.~\ref{Fig1}, panels (f) and (g).
The obtained skyrmion profile $\braket{\mathbf{s}}$ coincides with high accuracy with those obtained with 2D  DMRG simulations for square and disk geometries.
However, as expected, the amount of quantum fluctuations captured within the CMFA is smaller, leading to a weaker reduction of the magnetization length.
The crucial point is that the CMFA provides an lower bound on quantum fluctuations in the system and therefore represents a reliable and systematic first step from pure mean-field theory towards the study of quantum properties.

\textit{Quantum $k\pi$-Skyrmions} -- $k\pi$-skyrmions are examples of axially symmetric metastable states and are prime candidates for study within the CMFA.
For odd values of $k$, one can retain the same BCs as for the skyrmions above, while for even values of $k$ the classical spin at the origin has to be the same as on the circumference, i.e., $(0,0,s)$.
We focus on the simplest skyrmions of this type, which are stable at $h=0.65$ [see Fig.~\ref{Fig2}(a)].
Here, only $k=1$ and $k=3$ are topologically nontrivial ($Q=-1$) states, while the state with $k=2$, the so-called skyrmionium, is an example of a trivial state ($Q=0$).
The distribution of the magnetization length deviation $1-|\braket{\mathbf{s}_{i}}/s|$ is displayed in Fig.~\ref{Fig2}(b) and reaches maxima at the radius of the skyrmion (closest point to the origin with $m_{z}=0$)~\cite{Sallermann2021,Komineas2023}, and at the center of the rings ($m_{z}=-1$).
Surprisingly, among these three the most metastable state, the $3\pi$-skyrmion, shows the largest reduction in magnetization length with about $8\%$.

\textit{Quantum Hopfion rings} -- In the 3D case, axially symmetric $k\pi$-skyrmions can also be stabilized in the form of strings~\cite{Kuchkin2022} by tuning values of $h$ and $u$. 
One has to ensure $h+2u\geq 1$ to compensate for the spiral state along the $z$-axis and remain in the ferromagnetic phase.
A more non-trivial example of a state in such a background will be the hopfion ring around the skyrmion string.
Next, we study this state with the CMFA. The corresponding cluster is shown in Fig.~\ref{Fig3}(a).
Taking into account the axial symmetry, we use Eq.~\eqref{excpetc_values} to calculate all neighbors in the azimuthal directions, while for interactions along the $z$-axis we use the MFA $s_{ik}^{\alpha}s_{i,k\pm1}^{\beta}\to s_{ik}^{\alpha}\braket{s_{i,k\pm1}^{\beta}}$.
The BCs in the radial direction correspond to the skyrmion string state, while along the $z$-axis we impose periodic BCs.
First, we obtained stable skyrmion strings with zero, one, and two hopfion rings [Fig.~\ref{Fig3}(b)-(d)] by direct energy minimization of the classical model \eqref{E_dimensless} at $h=0.4$ and $u=0.3$.
They are all characterized by an identical topological charge $Q=-1$ but different hopfion charges~\eqref{Qhopf}, which equal the numbers of hopfion rings in the system~\cite{Azhar2024}.
Then we used these states as an initial guess for $\braket{\mathbf{s}_{ik}}$ in the CMFA simulations.

Within the CMFA, we obtain the spatial distribution of the magnetization length shown in Fig.~\ref{Fig3}(e)-(g).
For a pure skyrmion string (without hopfion ring), the maximum deviation of the magnetization length occurs at the skyrmion radius.
In contrast, in the presence of one (panel f) or two (panel g) hopfion rings, the regions of strongest quantum fluctuations are localized at the rings themselves.
Based on our simulations, the hopfion ring exhibits quantum fluctuations approximately three times larger than those in a pure skyrmion string.
It is noteworthy that the CMFA captures only a portion of the quantum fluctuations.
Simulations based on the full 3D Hamiltonian \eqref{XXZ} are therefore expected to reveal even stronger quantum effects. 
However, such calculations remain challenging at present and will motivate further research.

\textit{Zero mode of a hopfion rings} -- Based on symmetry considerations alone, the $z$ position of the hopfion ring is arbitrary on the skyrmion string. Therefore there is a corresponding translational zero mode for the hopfion ring.
To excite the zero mode, one can apply a small Zeeman field playing the role of a perturbation, i.e., $h\to h+\delta h$ with $\delta h\ll 1$.
In the MFA with $|\mathbf{m}|=1$, the classical Landau-Lifshitz (LL) equation, $\partial_{t}\mathbf{m}=\delta h \mathbf{m}\times\mathbf{e}_{z}$,
can also describe this zero mode~\cite{Kuchkin2025_v3}.
However, if one takes into account the quantum fluctuations which lead to regions with $|\mathbf{m}|\neq 1$, the LL equation predicts no motion of such quantum hopfion ring due to their conservation of magnetization length: $\partial_{t}|\mathbf{m}|^{2} = 2\mathbf{m}\cdot\partial_{t}\mathbf{m}=2\delta h\mathbf{m}\cdot\left(\mathbf{m}\times\mathbf{e}_{z}\right)=0$, because for hopfion ring motion, the regions with $|\mathbf{m}|\neq 1$ [see Fig.~\ref{Fig3}(f), (g)] must move as well. 

However, the recently introduced regularized LL equation~\cite{Kuchkin2026} for the motion of Bloch points can describe the zero mode for quantum hopfion rings.
Following the approach in Ref.~\cite{Kuchkin2025}, we introduce an $\mathbb{S}^{3}$-order parameter $\bm{\nu}$, with $|\bm{\nu}|=1$, which generalizes the magnetization vector, $\nu_{1}=m_{x}$, $\nu_{2}=m_{y}$, $\nu_{3}=m_{z}$, and $\nu_{4}^{2}=1-|\mathbf{m}|^{2}$, and by construction allows for the description of magnetization length dynamics while preserving the constraint $|\mathbf{m}|\leq 1$.
Without damping, the dynamics of the $\bm{\nu}$ components are given by
\begin{align}
\partial_{t}\nu_{1} &= \delta h \nu_{2}+\epsilon \delta h^{2}\nu_{1}\nu_{4}, & \partial_{t}\nu_{2} &= -\delta h \nu_{1}+\epsilon \delta h^{2}\nu_{2}\nu_{4}, \nonumber\\
\partial_{t}\nu_{3}&=0,& \partial_{t}\nu_{4} &= -\epsilon \delta h^{2}\left(\nu_{1}^{2} +\nu_{2}^{2}\right),\label{RLL}
\end{align}
where $\epsilon$ is a phenomenological parameter of the theory. For $\epsilon=0$, Eq.~\eqref{RLL} transforms into the standard LL equation.
In the case $\epsilon \neq 0$, the dynamical equations \eqref{RLL} allow for time-dependent magnetization length dynamics $\partial_{t}|\mathbf{m}|\neq 0$, and therefore they are compatible with the zero mode motion of a hopfion ring without excluding quantum fluctuations.

\textit{Conclusions} -- In this work, we applied the cluster mean-field approximation to study topologically nontrivial metastable states, $k\pi$-skyrmions and hopfion rings, in spin systems with a large number of spins. 
This approximation goes significantly beyond the mean field (classical) approximation and can effectively capture a portion of the quantum fluctuations present in the system.
We focused primarily on the fluctuations in the radial direction, motivated by the axial symmetry of the magnetic states of interest. 
In all cases, the CMFA revealed the existence of finite-size regions with varying magnetization length due to quantum spin entanglement.
Finally, we discussed the zero-mode dynamics of the hopfion ring along the skyrmion string and showed that the regularized Landau-Lifshitz equation can describe this motion, whereas the standard equation fails to capture the time-dependent magnetization length.

\textit{Acknowledgments} -- VMK is grateful to Nikolai Kiselev and \v{S}tefan Li\v{s}\v{c}\'{a}k for valuable discussions. 
VMK acknowledges the financial support from the European Union’s Horizon Europe research and innovation programme under the Marie Sk{\l}odowska-Curie grant agreement No.~101203692 (QUANTHOPF).
The authors acknowledge financial support from the Luxembourg National Research Fund under Grants C22/MS/17415246/DeQuSky and AFR/23/17951349.

\textit{Data Availability} -- The code for the DMRG simulations is based on the ITensor and ITensorMPS libraries~\cite{itensor}.
The code and raw data sets are published on Zenodo.

\newpage

\bibliography{main}

\newpage
\clearpage
\onecolumngrid

\begin{center}
   \textbf{Supplemental Material for ``Quantum Hopfion rings in the cluster mean-field approximation''}\newline 
\end{center}

\section{Hamiltonian dimensionalization}
\label{sec:Hamiltonian_dimless}

The micromagnetic form of a chiral magnet Hamiltonian is given by:
\begin{equation}
\mathcal{E}_{0}=\int_{\Omega}\mathrm{d}V\left[\mathcal{A}\left(\nabla\mathbf{m}\right)^{2}+\mathcal{D}\mathbf{m}\cdot\nabla\times\mathbf{m}+M_{s}B\left(1-m_{z}\right) + \mathcal{K}\left(1-m_{z}^{2}\right)\right],\label{micro}
\end{equation}
Exchange stiffness $\mathcal{A}$ and DMI constant $\mathcal{D}$ define the spin-spiral period $L_{D}=4\pi\mathcal{A}/\mathcal{D}$ and critical field $B_{D}=\mathcal{D}^{2}/2\mathcal{A}M_{s}$, which we employ to reduce the number of parameters in \eqref{micro} by introducing the magnetic field $h=B/B_\mathrm{D}$ and anisotropy $u=\mathcal{K}/B_\mathrm{D}$.
The dimensionless energy $\mathcal{E}=\mathcal{E}_{0}/\left(2\mathcal{A}L_\mathrm{D}\right)$ is provided in Eq.(2) in the main text.

As the Hamiltonian \eqref{micro} represents a continuum theory, it can be equivalently written in cylindrical coordinates $(r,\phi,z)$ using the following substitutions:
\begin{eqnarray}
&&\int_{\Omega} \mathrm{d}V=\int_{0}^{l}\mathrm{d}z\int_{0}^{R}r\mathrm{d}r\int_{0}^{2\pi}\mathrm{d}\phi,\label{cartesian_cylindrical}\\
&&\partial_{x}=\cos\phi\partial_{r}-\dfrac{\sin\phi}{r}\partial_{\phi},\quad \partial_{y}=\sin\phi\partial_{r}+\dfrac{\cos\phi}{r}\partial_{\phi},\nonumber
\end{eqnarray}
where the geometry of the magnetic sample is effectively a cylinder of radius $R$ and height $l$. 
Thus, the exchange and DMI energy densities are written as:
\begin{eqnarray}
    &&\!\!\!w_\mathrm{ex}(\mathbf{m})=\dfrac{1}{2}\left(\nabla\mathbf{m}\right)^{2}=\dfrac{1}{2}\left(\partial_{z} \mathbf{m}\right)^{2}+\dfrac{1}{2}\left(\partial_{r} \mathbf{m}\right)^{2} + \dfrac{1}{2r^{2}}\left(\partial_{\phi} \mathbf{m}\right)^{2},\label{w_ex_cylinder}\\
    &&\!\!\!w_\mathrm{dmi}(\mathbf{m})=\mathbf{m}\cdot\nabla\times\mathbf{m}=\mathbf{n}\cdot\nabla^{\prime}\times\mathbf{n}+\dfrac{n_{y}n_{z}}{r},\,\, \nabla^{\prime}=\left(\partial_{r},\dfrac{\partial_{\phi}}{r},\partial_{z}\!\right)\!,\label{w_dmi_cylinder}
\end{eqnarray}
where the field $\mathbf{n}$ has components $n_{x}=m_{x}\cos\phi+m_{y}\sin\phi$, $n_{y}=m_{y}\cos\phi-m_{x}\sin\phi$ and $n_{z}=m_{z}$.

\section{Quantum model in cylindrical coordinates}\label{sec:Quantum_Cylinder}

Utilizing \eqref{w_ex_cylinder} and \eqref{w_dmi_cylinder}, we can discretize the Hamiltonian Eq.(2) in the main text as follows:

\begin{eqnarray}
    &&\tilde{E}=\delta V\sum_{ijk}\left(r_{i+\frac{1}{2}}e_{r} + e_{\phi} + r_{i}e_{z} + 4\pi^{2}e_{u}\right),\label{XXZ_cylinder}\\
    &&e_{r}=\dfrac{1}{\delta r^{2}}-\dfrac{\mathbf{s}_{i+1,jk}\cdot\mathbf{s}_{ijk}}{s^2\delta r^{2}}+2\pi \dfrac{n_{i+1,jk}^{y*}s_{ijk}^{z}-s_{i+1,jk}^{z}n_{ijk}^{y*}}{s^{2}\delta r}\nonumber\\
    &&e_{\phi}=\dfrac{1}{r_{i}\delta \phi^{2}}-\dfrac{\mathbf{s}_{i,j+1,k}\cdot\mathbf{s}_{ijk}}{r_{i}s^2\delta \phi^{2}}+2\pi \dfrac{n_{ijk}^{x*}s_{i,j+1,k}^{z}-s_{ijk}^{z}n_{i,j+1,k}^{x*}}{s^{2}\delta\phi},\nonumber\\
    &&e_{z}=\dfrac{1}{\delta z^{2}}-\dfrac{\mathbf{s}_{ij,k+1}\cdot\mathbf{s}_{ijk}}{s^2\delta z^{2}}+2\pi\dfrac{s_{ij,k+1}^{x}s_{ijk}^{y}-s_{ij,k+1}^{y}s_{ijk}^{x}}{s^{2}\delta z},\nonumber\\
    &&e_{u}=r_{i}\left(h+2u\dfrac{\braket{s_{ijk}^{z}}}{s}\right)\left(1-\dfrac{s_{ijk}^{z}}{s}\right),\nonumber
\end{eqnarray}
where $n^{x*}=s^{x}\cos\phi_{j+\frac{1}{2}}+s^{y}\sin\phi_{j+\frac{1}{2}}$ and $n^{y*}=s^{y}\cos\phi_{j}-s^{x}\sin\phi_{j}$ for all sets of indices present in expressions $e_{\phi}$ and $e_{r}$, respectively.
The accuracy of the discretization depends on the smallness of
$\delta z=1/N_{z}$, and $\delta r=1/N_{r}$ as analog to $a$ but in the radial direction, and $\delta \phi=2\pi/N_{\phi}$ in azimuthal direction, where the total number of spins is given by $N_{r}N_{\phi}N_{z}$.
The volume element in \eqref{XXZ_cylinder} is denoted as $\delta V=\delta r \delta \phi \delta z$.

The Hamiltonian \eqref{XXZ_cylinder} is used to perform CMF simulations.
In particular for axially symmetric 2D skyrmions, we have the following effective 1D Hamiltonian for each cluster:
\begin{eqnarray}
    &&E_\mathrm{1D}=\delta V\sum_{i=0}^{N_{r}}\left(r_{i+\frac{1}{2}}e_{r}^{\prime} + e_{\phi}^{\prime} + 4\pi^{2} e_{u}^{\prime}\right),\label{E_1D}\\
    &&e_{r}^{\prime}=\dfrac{1}{\delta r^{2}}-\dfrac{\mathbf{s}_{i+1}\cdot\mathbf{s}_{i}}{s^2\delta r^{2}}+2\pi \dfrac{n_{i+1}^{y*}s_{i}^{z}-s_{i+1}^{z}n_{i}^{y*}}{s^{2}\delta r}\nonumber\\
    &&e_{\phi}^{\prime}=\dfrac{2}{r_{i}\delta\phi^{2}}-2\dfrac{\left(s_{i}^{x}\braket{ s_{i}^{x}} +s_{i}^{y}\braket{s_{i}^{y}} \right)\cos\delta\phi+s_{i}^{z}\braket{ s_{i}^{z}} }{r_{i}s^2\delta \phi^{2}}+4\pi \dfrac{\left(\!\braket{s_{i}^{z}}\! s_{i}^{y}\!+\!s_{i}^{z}\braket{s_{i}^{y}}\! \right)\cos\phi^{*}\!+\!\left(\!\braket{s_{i}^{z}}\!s_{i}^{x}\!-\!s_{i}^{z}\!\braket{s_{i}^{x}}\!\right)\sin\phi^{*}}{s^{2}\delta\phi}\sin\dfrac{\delta\phi}{2},\nonumber\\
    &&e_{u}^{\prime}=r_{i}\left(h+2u\dfrac{\braket{s_{i}^{z}}}{s}\right)\left(1-\dfrac{s_{i}^{z}}{s}\right).\nonumber
\end{eqnarray}
Here, index $j$ denotes the cluster, $\phi^{*}=j \delta \phi$.

In a 3D case, we have an effective 2D Hamiltonian:
\begin{eqnarray}
    &&E_\mathrm{2D}=\delta V\!\sum_{ik}\!\left(r_{i+\frac{1}{2}}e_{r}^{\prime} + e_{\phi}^{\prime} + r_{i}e_{z}^{\prime} + 4\pi^{2}e_{u}^{\prime}\!\right),\!\label{E_2D}\\
    &&e_{z}^{\prime}=\dfrac{2}{\delta z^{2}}-\dfrac{\mathbf{s}_{ik}\cdot\left(\braket{\mathbf{s}_{i,k-1}} + \braket{\mathbf{s}_{i,k+1}}\right)}{s^2\delta z^{2}}+2\pi\dfrac{\braket{s_{i,k+1}^{x}}s_{ik}^{y}-\braket{s_{i,k+1}^{y}}s_{ik}^{x} + s_{ik}^{x}\braket{s_{i,k-1}^{y}}-s_{ik}^{y}\braket{s_{i,k-1}^{x}}}{s^{2}\delta z},\nonumber
\end{eqnarray}
where expressions for $e_{r}^{\prime}$, $e_{\phi}^{\prime}$ and $e_{u}^{\prime}$ are provided in \eqref{E_1D}.

\section{Convergence of 2D DMRG simulations}
\label{sec:DMRG_convergnece}

\begin{figure*}[tb!]
\centering
\includegraphics[width=17.7cm]{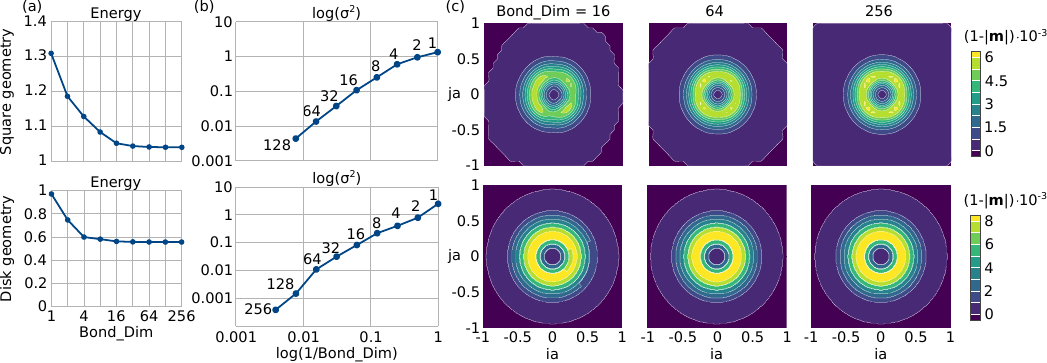}
\caption{~\small \textbf{Convergence analysis of 2D DMRG simulations for a skyrmion}. 
Upper (lower) panel (a) skyrmion energy and (b) variance, \eqref{sigma2}  for skyrmion as a function of bond dimensions in square (disk) geometry.
(c) corresponding magnetization length distribution for selected bond dimensions $16$, $64$ and $256$.
} 
\label{FigA1}
\end{figure*}

As the DMRG method relies on the matrix product state (MPS) form of the wave function, which has been theoretically proven to work in the 1D case~\cite{Hastings2007}, it is always important to examine its convergence in higher dimensions to ensure the results are trustworthy. 
The bond dimension is a parameter responsible for the accuracy of this representation, and we systematically examined its role in DMRG simulations of 2D models \eqref{XXZ} and \eqref{XXZ_cylinder}.
As shown in Fig.~\ref{FigA1}(a), the energies of the skyrmion states are saturated for bond dimensions $\gtrsim 16$.
The exact energy values for the square Eq.(3) in the main text and disk \eqref{XXZ_cylinder} geometries do not match because we used different numbers of particles (960 for the square and 256 for the disk), and the contribution of quantum fluctuations to the energy differs significantly due to the different underlying Hamiltonian forms.
Another quantity that one can look at is the variance, defined for a quantum Hamiltonian $H_{q}$ and for a state with energy $E_{q}$ as:
\begin{eqnarray}
    \sigma^{2}=\braket{\left(H_{q}-E_{q}\right)^{2}},\label{sigma2}
\end{eqnarray}
which is shown in Fig.~\ref{FigA1}(b). 
$\sigma^{2}$ typically decreases as the bond dimension increases~\cite{Haller2022,Lik2026}.
The values of $\sigma^{2}$ are smaller in the case of disk geometry for a fixed bond dimension, which can be explained by a different number of particles used in both cases.
The distribution of $1-|\mathbf{m}|$ becomes more axially symmetric for higher bond dimensions, as demonstrated in Fig.~\ref{FigA1}(c). 
Thus, a symmetry of observables can serve as an additional benchmark for the DMRG simulation results, alongside energy saturation and variance zeroing.

\end{document}